**[Article Full Title]**

Artificial Intelligence-Facilitated Online Adaptive Proton Therapy Using Pencil Beam Scanning Proton Therapy

**[Short Running Title]**

Prostate oAPT in PBSPT

**[Author Names]**

Hongying Feng, PhD[2,1,3], Jie Shan, MS[1], Carlos E. Vargas, MD[1], Sameer R. Keole, MD[1], Jean-Claude M. Rwigema, MD[1], Nathan Y. Yu, MD[1], Yuzhen Ding, PhD[1], Lian Zhang, PhD[1], Steven E. Schild, MD[1], William W. Wong, MD[1], Sujay A. Vora, MD[1], JiaJian Shen, PhD[1], Wei Liu, PhD[1]

**[Author Institutions]**

[1]Department of Radiation Oncology, Mayo Clinic, Phoenix, AZ 85054, USA

[2]College of Mechanical and Power Engineering, China Three Gorges University, Yichang, Hubei 443002, China

[3]Department of Radiation Oncology, Guangzhou Concord Cancer Center, Guangzhou, Guangdong, 510555, China

**[Corresponding Author Name & Email Address]**

Wei Liu, PhD, e-mail: Liu.Wei@mayo.edu.

**[Author Responsible for Statistical Analysis Name & Email Address]**

Wei Liu, PhD, e-mail: Liu.Wei@mayo.edu.

**[Conflict of Interest Statement for All Authors]**

None

**[Funding Statement]**

This research was supported by *Arizona Biomedical Research Commission Investigator Award (R01 type)*, by the *President's Discovery Translational Program of Mayo Clinic*, by the *Fred C. and Katherine B. Anderson Foundation Translational Cancer Research Award*, by The *Lawrence W. and Marilyn W. Matteson Fund for Cancer Research*, and by The *Kemper Marley Foundation.*

**[Data Availability Statement for this Work]**

Research data are stored in an institutional repository and will be shared upon request to the corresponding author

**[Acknowledgements]**

None

**Abstract**

**Purpose:** Online adaptive proton therapy (oAPT) is essential to address interfractional anatomical changes in patients receiving pencil beam scanning proton therapy (PBSPT). Artificial intelligence (AI)-based auto-segmentation can increase the efficiency and accuracy. Linear energy transfer (LET)-based biological effect evaluation can potentially mitigate possible adverse events caused by high LET. New spot arrangement based on the verification CT (*vCT*) can further improve the re-plan quality. We propose an oAPT workflow that incorporates all these functionalities and validate its clinical implementation feasibility with prostate patients.

**Methods and Materials:** AI-based auto-segmentation tool AccuContour[TM] (Manteia, Xiamen, China) was seamlessly integrated into oAPT. Initial spot arrangement tool on the *vCT* for re-optimization was implemented using raytracing. An LET-based biological effect evaluation tool was developed to assess the overlap region of high dose and high LET in selected OARs. Eleven prostate cancer patients were retrospectively selected to verify the efficacy and efficiency of the proposed oAPT workflow. The time cost of each component in the workflow was recorded for analysis.

**Results:** The verification plan showed significant degradation of the CTV coverage and rectum and bladder sparing due to the interfractional anatomical changes. Re-optimization on the *vCT* resulted in great improvement of the plan quality. No overlap regions of high dose and high LET distributions were observed in bladder or rectum in re-plans. 3D Gamma analyses in PSQA confirmed the accuracy of the re-plan doses before delivery (Gamma passing rate = 99.57 ± 0.46%), and after delivery (98.59 ± 1.29%). The robustness of the re-plans passed all clinical requirements. The average time for the complete execution of the workflow was 9.12 ± 0.85 minutes, excluding manual intervention time.

**Conclusion:** The AI-facilitated oAPT workflow was demonstrated to be both efficient and effective by generating a re-plan that significantly improved the plan quality in prostate cancer treated with PBSPT.Keywords: online adaptive radiation therapy, artificial intelligence-based auto-segmentation, pencil beam scanning proton therapy, robust optimization, prostate cancer4

**Introduction**

Pencil beam scanning proton therapy (PBSPT) is the most advanced form of proton therapy, featured by the high flexibility at the beamlet level in treatment planning and dose delivery, leading to superior performance in target coverage and organs at risk (OARs) protection[1-3] compared to photon therapy or passive scattering proton therapy. Though precise, PBSPT is highly sensitive to uncertainties caused by many factors, including proton beam range and patient setup uncertainties, respiratory motion, and anatomical changes[4,5]. Robust optimization has been proposed to account for patient setup, range uncertainties, and respiratory motion in PBSPT[6-17]. To address the anatomical changes, adaptive radiation therapy (ART) has been introduced[18,19]. ART involves undergo periodic verification imaging during treatment course to obtain information about the internal anatomical changes of the patient. The clinician will assess whether the dose from the initial plan is still within the allowable tolerance based on the periodic verification images, and re-planning would be needed if the initial plan does not meet clinical requirements.

For ART, the online ART (adaption just before delivery during one fraction) outperformed the offline ART (adaption between fractions) in terms of target coverage and normal tissue protection owing to minimization of critical time delays. Nonetheless, successful implementation of the online ART, which usually take 45 to 80 minutes for Halcyon™ (Varian Medical Systems, Palo Alto, CA), MRIdian™ (ViewRay, Oakwood Village, OH), and Unity™ (Elekta, Stockholm, Sweden) in photon therapy[20-23], poses a much higher demand on the efficiencies of each component of ART (image acquisition, organ segmentation, dose verification, re-optimization, and patient-specific quality assurance (PSQA)) as well as their integration as a whole.

In PBSPT, multiple reports have been published about online adaptive proton therapy (oAPT), including dose restoration-based methods[24,25], deformable image registration (DIR)-based beamlet position/energy adaption, and weight tunning[26], and rigid image registration (RIR)-based structure propagation and robust re-optimization[27]. All of these are applicable for some clinical scenarios with small anatomically changes. Investigators recently proposed a Monte Carlo (MC)-based oAPT workflow that was applicable to large anatomical changes taking about 15 minutes to be finished, excluding possible manual intervention powered by graphic processing unit (GPU) acceleration[28]. However, the proposed workflow also faced several challenges: (1) the accuracy of the DIR-based auto-segmentation heavily depends on the accuracies of the reference contours on the initial planning CT and the DIR algorithm, and (2) the commercial treatment planning system, Eclipse$^{TM}$ (Varian Medical System, Palo Alto, CA), has to be used for new initial spot arrangement based on the verification CT with large anatomical changes.

In addition, the generally accepted constant relative biological effectiveness (RBE) of 1.1 in proton therapy underestimates RBE at the Bragg Peaks, which exhibits high linear energy transfer (LET)[29]. The underestimation of RBE can potentially result in overshooting of biological dose, increasing the risk of adverse events[29,30]. Therefore LET, as a surrogate of RBE, is increasingly utilized in plan evaluation and even in treatment planning in proton therapy[12,31-33].

In this work, our aim is to meet the aforementioned challenges via: (1) integrating the artificial intelligence (AI)-based auto-segmentation tool, AccuContour$^{TM}$ (Manteia, Xiamen, China), to delineate OARs from scratch and circumvent the segmentation accuracy dependent on contour propagation using DIR algorithms, (2) developing an initial spot arrangement tool by raytracing to get an optimal initial spot arrangement based on the verification CT (*vCT*), and (3) developing an LET-based biological effect evaluation tool to enable the direct quantitative assessment of the LET distribution in OARs.

In addition, in this study we evaluated a cohort of patients with treated with PBSPT to investigate the feasibility of the proposed oAPT workflow in clinical implementation, because of two reasons: (1) prostate cancer is relatively easy to treat, thus it is an ideal disease site to clinically implement such a complicated and brand new treatment technique as oAPT for the first time and (2) due to the daily anatomy changes resulting from rectum distention, inconsistent bladder filling, and frequent bowel gas, etc., oAPT would better benefit prostate cancer treatment with PBSPT compared to the offline ART.

**Methods and Materials**

**A. Artificial Intelligence-Facilitated Online Proton Adaptive Workflow**

The oAPT workflow generally consists of four steps[28], including preparation, verification, re-optimization, and patient-specific quality assurance (PSQA). In the preparation step, all the patient anatomy information (CT and structure DICOMs) and treatment plan information (plan and dose DICOMs) needed for the following steps are prepared. In the verification step, a verification plan is generated by performing a forward dose calculation on *vCT* using the initial plan. In the re-optimization step, a robust re-optimization with the generated new structure set will be carried out on the *vCT* to generate a re-plan if the verification plan fails to meet the clinical requirements. In the PSQA step, a two-stage PSQA is done. For stage 1 before-delivery, a second dose of the re-plan was calculated using an independent MC dose engine. For stage 2 after-delivery, a delivered plan reconstruction and dose calculation using the independent MC dose engine were performed based on log files collected during treatment delivery. Please refer to XXXX et al.[28] for more details. We have introduced new functionalities into the above general clinical workflow as outlined below.

**A.1 Integrating AccuContour$^{TM}$ for Auto-Segmentation of Organs at Risk**

AccuContour$^{TM}$ is a commercial AI software for auto-segmentation in radiation therapy. AccuContour$^{TM}$ provides a collection of pre-trained models for different disease sites[34,35]. Besides, AccuContour$^{TM}$ also provides an application programming interface (API), that can be incorporated in a third-party application.

We have developed an in-house treatment planning system (TPS), named XXXX[9,14,17,36,37], to serve as the foundation for the oAPT workflow development, as depicted in Figure 1 (shapes with solid edge lines). The client, featuring a Window GUI and running on a floating workstation, is seamlessly implemented as an Eclipse™ (Varian Medical System, Palo Alto, CA) plugin via Eclipse Scripting API (ESAPI), while the server, running on a dedicated Linux server, communicates with the client via transmission control protocol (TCP) socket communication. The server retrieves and sends DICOM files with the DB Daemon DICOM server via the pynetdicom Python package.

The shapes with dashed edge lines in Figure 1 demonstrated the newly added parts related to AccuContour™. The AccuContour™ API is integrated into the client, giving commands to and listening to the callback status from the AccuContour™ server. The commands dictate three specific tasks: (1) retrieving DICOMs from the DB Daemon DICOM server to the AccuContour™ server for auto-segmentation, (2) delineating the user-selected structures with the user-customized AI models within the AccuContour™ server, and (3) sending the generated structure DICOM files back to the DB Daemon DICOM server. The callback function informs the AccuContour™ API the status of a specific task on the AccuContour™ server. The AccuContour™ server communicates with the DB Daemon DICOM server via DCMTK (OFFIS, Oldenburg, Germany) for DICOM file retrieving.

Notably, the integrated AccuContour™ is still restricted to the auto-segmentation tasks for OARs, though much effort has been dedicated to exploit its ability in target delineation[38,39]. Therefore, a RIR was additionally introduced to propagate all the targets from the initial planning CT(*iCT*) to *vCT*. The propagated targets were then passed to the AccuContour™ server to be combined with the AI-segmented OARs for the generation of the new structure DICOM file on the *vCT* (Fig. 1). The newly generated structure DICOM file requires the editing and approval from physicians, facilitated by physicists and dosimetrists.

In this study, prostate, bladder, rectum, left and right femoral heads, penile bulb, seminal vesical, and rectum spacer were selected to form a template for the AI-based auto-segmentation in the proposed oAPT workflow.

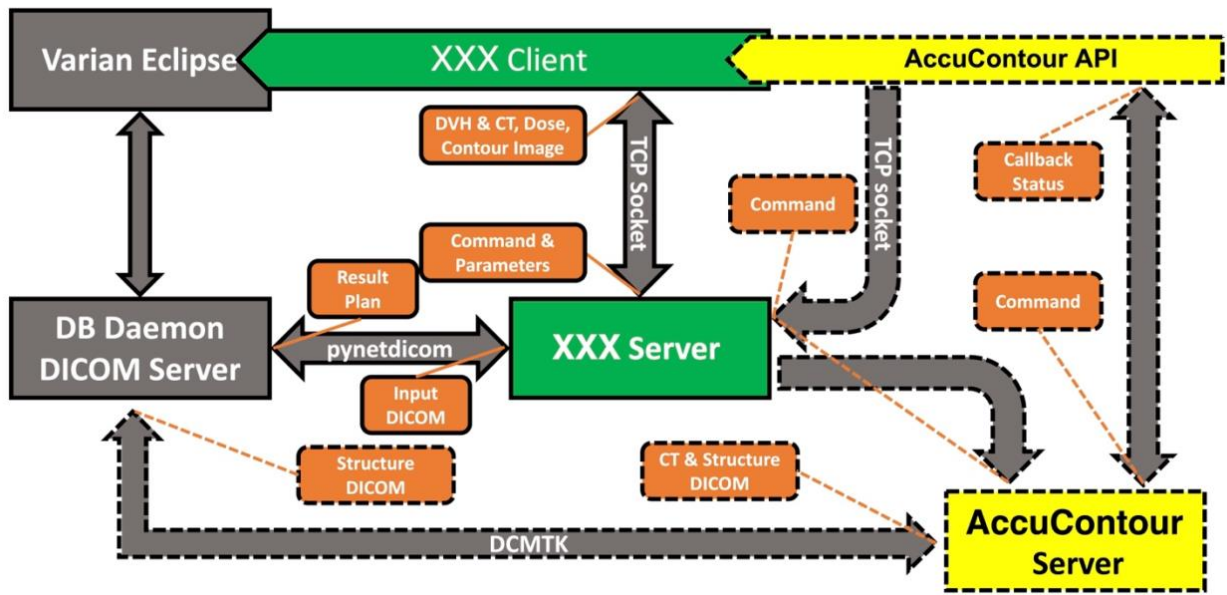

Figure. 1 Pipeline of the enhanced oAPT workflow for auto-segmentation. The shapes with solid edge lines are the prototype pipelines, while shapes with dashed edge lines are the newly added pipelines related to AccuContour$^{TM}$.

**A.2 Initial Spot Arrangements through Raytracing for Re-Optimization on *vCT***

Once the new targets for initial spot arrangement were approved (gray boxes in Fig 2[a]) and the isocenter was determined (via a rigid registration), a 40 cm X 30 cm field, in which the spots were uniformly distributed in either a hexagonal pattern or a square pattern with a spot spacing of 5 mm, was prepared for the initial spot arrangement. In treatment planning for prostate patients, optimization target volume (OTV) was constructed by adding a 3 mm margin to the clinical target volume (CTV) in all directions except 8 mm in lateral directions to include range uncertainty per our institution protocol. Usually a scanning target volume (STV) was formed by expanding optimization target volume (OTV) for initial spot arrangement to make sure that there was at least one spot outside OTV to guarantee that a uniform dose distribution could be formed within OTV[40]. For a certain lateral position from a certain beam angle, the 97 discrete proton energies from 71.3 MeV to 228.8 MeV used clinically at our proton center, were scanned to determine whether a certain proton energy should be kept or not based on the water equivalent thickness (WET) calculated by raytracing along the beam direction from the virtual radiation source to the specified voxel. Figure 2(a) gives a conceptional demonstration of the methodology. The algorithm to determine whether a certain proton energy was kept is illustrated in Figure 2(b).

Notably, for the different treatment fields in a plan, the STV used by each treatment field for initial spot arrangement does not necessarily to be the same, i.e., field-specific STV is supported.

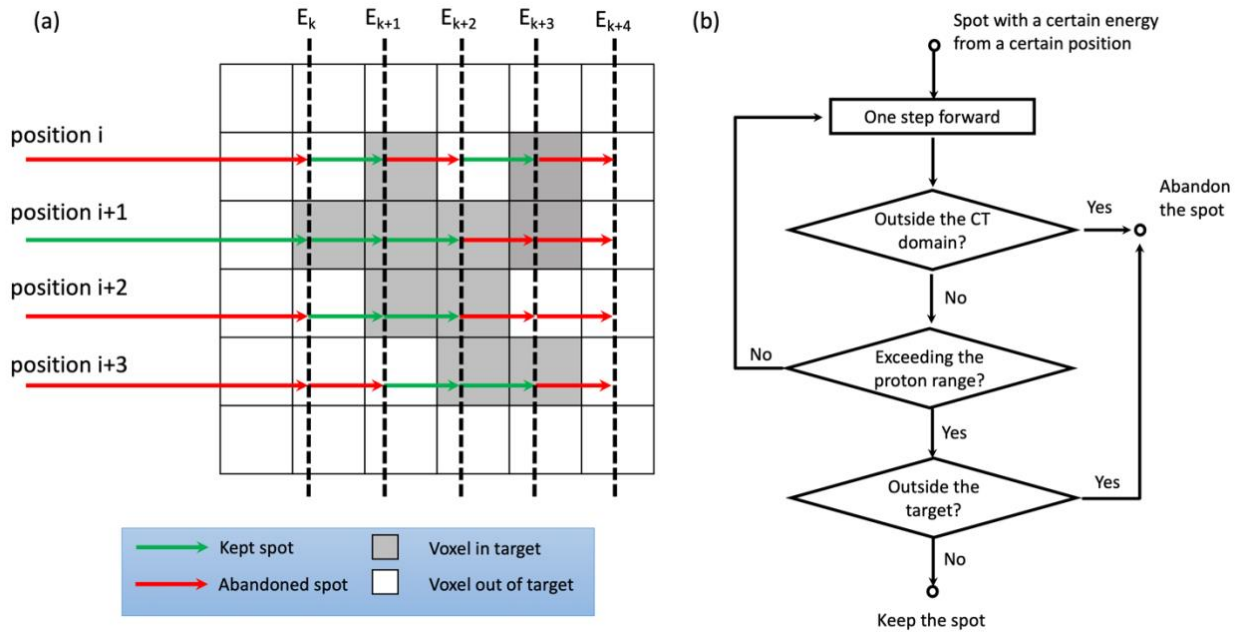

Figure. 2 Sketch of the raytracing method for initial spot arrangement. (a) A 2D conceptual demonstration of raytracing. For each position (from i to i+3), proton energy levels (from k to k+4) are scanned. The dashed vertical lines indicate the stopping positions of spots with the corresponding proton energies annotated above. The green arrows indicate the kept proton energies, while the red arrows denote the left-out proton energies. (b) The loop to determine whether a spot of a certain proton energy should be kept or left out.

### A.3 LET-based Biological Effect Evaluation Tool

When an MC-based verification plan dose calculation was finished, the resultant dose and LET distributions were jointly used for the LET-based biological effect evaluation. Users first chose the structures of interest (SOIs) and then set up the SOI-specific dose and LET thresholds. By applying the dose and LET thresholds to the SOI dose and LET distributions correspondingly, the overlap region between the high dose region (beyond the dose threshold) and high LET region (beyond the high LET threshold) within the SOI could be derived. Such overlap regions have increasingly been recognized to increase the risk of adverse events[33,41,42] and should be minimized.

In this study, the 50% of the prescribed dose and 6 keV/um were selected as the dose and LET thresholds respectively per our institutional protocol[43].

**B. Patient Selection**

Eleven patients with prostate cancer treated with IMPT at our institution from March 2020 to December 2022 were retrospectively selected for this study with the following inclusion criteria: (1) all patients had the same prescription dose (70 Gy[RBE] delivered in 28 fractions); (2) all patients had the same treatment field configuration (2 slightly backward rotated lateral beams, i.e., T180G95 and T0G95 with T for the table angle and G for the gantry angle)[42]; and (3) for each patient at least one re-plan was introduced during the treatment course.

**C. Verification Plan and Re-Plan**

The initial plan was evaluated on the *vCT*, which was acquired by CT-on-rails (Siemens, Erlangen, Germany). A RIR was first done between the *vCT* and *iCT* by Elastix[44], with *vCT* being the reference image. Based on the RIR results, the original plan was transformed to the *vCT* to conduct a forward dose calculation on the *vCT* to generate the verification plan. Then, with the AI-based auto-segmented OARs and rigidly propagated targets (the same density override as in the initial plan would be performed automatically based on the newly generated structure set), a dosimetrical evaluation of the verification plan was carried out using ClearCheck™ (Radformation, New York, NY) per our institutional protocol of the dose volume histogram (DVH) indices (Table 1). For some constraints, a range rather than an exactly number was used, with the left limit as soft limit that better be satisfied and the right limit as hard limit that must be met.

Table 1. Dose-volume constraints (DVCs) for prostate patients treated with IMPT.

| Structure | Dose Volume Constraints (DVCs) |
|---|---|
| OTV | $D_{max} < 107\text{-}110\%$ |
| CTV | $V_{98\%} > 95\text{-}94\%$; $V_{100\%} > 98\%$ |
| Bladder | $V_{39Gy[RBE]} < 50\%$; $V_{44Gy[RBE]} < 40\text{-}50\%$; $V_{55Gy[RBE]} < 25\text{-}30\%$; $V_{61Gy[RBE]} < 20\text{-}25\%$; $V_{66Gy[RBE]} < 15\text{-}20\%$ |
| Rectum | $V_{35Gy[RBE]} < 40\text{-}50\%$; $V_{53Gy[RBE]} < 20\text{-}30\%$; $V_{61Gy[RBE]} < 15\text{-}20\%$; $V_{66Gy[RBE]} < 10\text{-}15\%$; $V_{105\%} < 1cc$ |

*abbreviations*: OTV for optimization target volume, CTV for clinical target volume, RBE for relative biological effectiveness (RBE = 1.1).

A re-plan was triggered when the verification plan failed to meet the clinical requirements. At first, STV was selected for the initial spot arrangement as described in section A.2. Then a robust re-optimization to the CTV was done based on the set of the newly generated spots using VPMC. In the robust re-optimization, proton range uncertainties were considered by scaling the nominal proton range up and down by 3%. As the robust re-optimization in the proposed oAPT workflow was directly carried out on the CT-of-the-day with the patient still fixed on the treatment table, the patient setup uncertainties were expected to be small, thus a 1 mm rigid shift (instead of the routinely used 3 mm) in all six directions was used to addressed the patient setup uncertainties[27]. Altogether, a set of 13 different uncertainty scenarios was used (one nominal and 12 perturbed scenarios). Single-field optimization (SFO) was used in the initial plan optimization, which was also applied for the re-optimization.

## D. Plan Robustness Evaluation

A separate plan robustness evaluation procedure was carried out using VPMC with the same uncertainty scenario settings as used in the re-optimization. The DVH indices in the worst-case scenario were used to indicate the plan robustness. For the target (CTV), $D_{95\%}$ of the CTV in the worst-case scenario was required to be higher than 99% of the prescription dose when treating prostate patients per our institution protocol, while for the important OARs (rectum and bladder), one must make sure that dose distributions in the worst-case scenario should still meet the clinical requirements as shown in Table 1.

## Results

### A. Plan Evaluation

The initial plan, verification plan, and re-plan were compared using ClearCheck[TM]. Four typical clinically relevant DVH indices that best demonstrate the efficacy of the oAPT workflow was selected from what were listed in Table 1 and shown in Figure 3, for the initial plan, verification plan, and re-plan respectively. Other DVH indices were shown in Figure S1 in the supplementary materials.

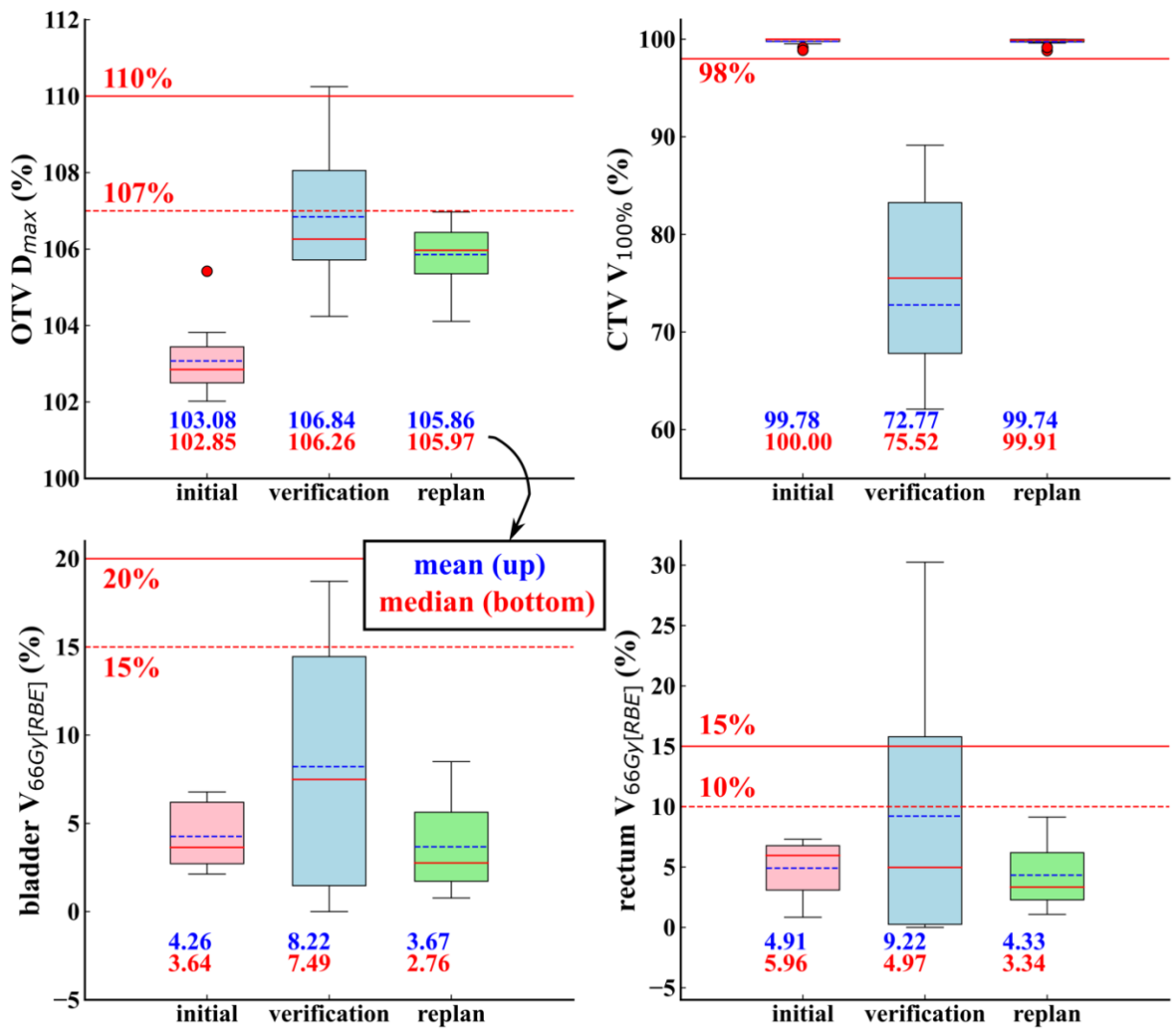

Figure. 3 Comparison of the dose-volume histogram indices among the initial plans (pink), verification plans (light blue), and re-plans (light green). Horizontal red lines in each subpanel

indicate our institutional dose volume constraints, with solid line being the hard limit that must be met and dashed line being the soft limit that would better be met.

For the targets, all the initial plans satisfied all the clinical dose volume constraints (DVCs) for OTV $D_{max}$ (median: 102.85%) and CTV $V_{100\%}$ (median: 100.00%). However, in the verification plans, for the OTV $D_{max}$, 1 of 11 patient did not pass the hard constraint, while 2 patients did not pass the soft constraint; for the CTV $V_{100\%}$, all the patients failed to meet the hard constraint. After the re-optimization, all the re-plans satisfied all the clinical DVCs for OTV $D_{max}$ (median: 105.97%) and CTV $V_{100\%}$ (median: 99.91%). For the OARs, all the initial plans met every DVC for both bladder and rectum. However for bladder, the verification plan from two patient did not pass the soft limits for $V_{66Gy[RBE]}$. For rectum, three patients failed to meet the hard limits for $V_{66Gy[RBE]}$. After the re-optimization, the plan quality was improved that all the re-plans met every DVC for both bladder and rectum.

As for the LET-based biological effect evaluation, when the default thresholds of dose (50% of prescription) and LET (6 keV/um) were used, all re-plans passed the LET-based biological effect evaluation for bladder and rectum, i.e., no overlap regions of high dose and high LET distributions (above the thresholds respectively) were observed.

**B. Re-plan Robustness Evaluation**

The re-plans of 10 of 11 patients met all of the clinicaly relevant DVCs , except that the re-plan of the other one patient slightly failed the soft constraints for rectum $V_{61Gy[RBE]}$ and $V_{66Gy[RBE]}$ (see Table S1 in the supplementary material). Nonetheless, the plan robustness for this patient was still considered acceptable since the hard constraints were strictly satisfied.

## C. Re-plan PSQA

The 3D Gamma passing rates for PSQA before-delivery PSQA and after-delivery PSQA with a threshold of 2%/2 mm/10% are shown in Table 2. For the before-delivery PSQA, the 3D Gamma passing rate was 99.57 ± 0.46%. For the after-delivery PSQA, the corresponding 3D Gamma passing rate was 98.59 ± 1.29%. The high 3D Gamma passing rates verified the accuracy of the re-plan dose using an independent Monte Carlo dose engine before-delivery (2nd dose check) and after-delivery (using the log files generated during the plan delivery) PSQA.

Table. 2 3D Gamma passing rates with a threshold of 2%/2 mm/10% for both before- and after-delivery PSQA of all patients

| Patiant (#) | Before-delivery | After-delivery |
|---|---|---|
| 1 | 99.83% | 99.52% |
| 2 | 99.80% | 99.27% |
| 3 | 99.72% | 99.47% |
| 4 | 99.75% | 99.65% |
| 5 | 99.76% | 99.32% |
| 6 | 99.68% | 99.45% |
| 7 | 99.36% | 95.33% |
| 8 | 98.18% | 98.55% |
| 9 | 99.74% | 97.17% |
| 10 | 99.81% | 99.15% |
| 11 | 99.62% | 97.61% |
| Ave. | 99.57% | 98.59% |
| Std. | 0.46% | 1.29% |

## D. Time Cost of the oAPT workflow

The time cost for each component of the proposed oAPT workflow was shown in Table 3. Human intervention that was required after the generation of the new structure DICOM file for manual editing and approval was excluded from the time cost since it was highly patient and user specific. The after-delivery log file-based PSQA was also excluded because such step did not have to be carried out during the oAPT workflow. Rather, it could be executed anytime between the oAPT workflow for the current fraction and the delivery of the next fraction. The image registration step consisted of two substeps, CT DICOM files transfer and RIR (in italic in Table 4). The AI-based new structure DICOM file generation step consisted of three substeps, transfering CT DICOM files and structure DICOM file to the AccuContour$^{TM}$ server, auto-delineation, and transfering the structure DICOM file back to the DB Deamon DICOM server (in italic in Table 4). In total, the average time cost was 547.3 $\pm$ 51.1 seconds, with AI-based new structure DICOM file generation (164.2 $\pm$ 26.0 seconds), re-optimization (155.5 $\pm$ 40.8 seconds), and image registration (84.1 $\pm$ 6.6 seconds) as the three most time-consuming components.

Table 3 Time cost for each component of the proposed oAPT workflow (in seconds).

| Patient # | 1 | 2 | 3 | 4 | 5 | 6 | 7 | 8 | 9 | 10 | 11 | Ave. | Std. |
|---|---|---|---|---|---|---|---|---|---|---|---|---|---|
| Image Registration | 84 | 91 | 83 | 95 | 88 | 92 | 77 | 76 | 83 | 77 | 79 | 84.1 | 6.6 |
| *DICOM transfer* | *54* | *62* | *58* | *68* | *63* | *68* | *51* | *52* | *58* | *51* | *53* | *58.0* | *6.1* |
| *Rigid registration* | *30* | *29* | *25* | *27* | *25* | *24* | *26* | *24* | *25* | *26* | *26* | *26.1* | *1.8* |
| New structure DICOM generation (rigid) | 24 | 26 | 31 | 29 | 30 | 25 | 23 | 32 | 27 | 25 | 29 | 27.4 | 2.9 |
| New structure DICOM generation (AI) | 157 | 183 | 202 | 214 | 174 | 130 | 125 | 148 | 172 | 161 | 161 | 164.2 | 26.0 |
| *DICOM transfer 1*[a] | *17* | *19* | *22* | *22* | *18* | *14* | *16* | *15* | *17* | *16* | *15* | *17.4* | *2.6* |
| *Auto-Contour* | *125* | *146* | *160* | *171* | *139* | *102* | *97* | *117* | *140* | *130* | *128* | *130.3* | *21.4* |
| *DICOM transfer 2*[b] | *15* | *18* | *20* | *21* | *17* | *14* | *12* | *16* | *15* | *15* | *18* | *16.5* | *2.5* |
| Verification dose evaluation | 24 | 23 | 26 | 24 | 23 | 18 | 19 | 21 | 27 | 28 | 33 | 24.2 | 4.1 |
| ClearCheck evaluation | 14 | 13 | 16 | 15 | 15 | 19 | 13 | 14 | 15 | 13 | 15 | 14.7 | 1.7 |
| Initial spot generation and IM calculation | 36 | 41 | 42 | 26 | 28 | 18 | 17 | 19 | 20 | 22 | 25 | 26.7 | 8.7 |
| Re-optimization | 216 | 89 | 168 | 141 | 131 | 161 | 188 | 187 | 96 | 125 | 208 | 155.5 | 40.8 |
| Before-delivery PSQA | 29 | 34 | 32 | 27 | 30 | 20 | 23 | 26 | 21 | 19 | 24 | 25.9 | 4.8 |
| Robust evaluation | 29 | 35 | 33 | 22 | 23 | 16 | 16 | 17 | 18 | 20 | 21 | 22.7 | 6.4 |
| Total | 613 | 535 | 633 | 593 | 542 | 499 | 501 | 540 | 479 | 490 | 595 | 547.3 | 51.1 |

*abbreviations*: AI for artificial intelligence, IM for influence matrix, PSQA for patient specific quality assurance.

[a] Transfering CT DICOM files and structure DICOM file to the AccuContour<sup>TM</sup> server.

[b] Transfering the structure DICOM to the DB Deamon DICOM server.

**Dicussion**

In this study, we have successfully incoporated new functionalities into our in-house TPS to implement AI-facilitated MC-based oAPT workflow for PBSPT: (1) an AI-based auto-segmentation tool, AccuContour$^{TM}$, for auto-segmentation of OARs, (2) an initial spot arrangement tool by raytracing to get optimal initial spots on *vCT*, and (3) an LET-based biological effect evaluation tool to enable the direct quantitative assessment of the LET distribution. Eleven prostate patients were retrospectively selected to validate and demonstrate the effectiveness of the proposed oAPT workflow. With the selected patient cohort, the average time cost to conduct the whole oAPT workflow was 547.3 ± 51.1 seconds, excluding human intervine in manual editing and approval of the newly generated structure set.

In this work, we selected prostate cancer patients as a first attempt to investigate the feasibility of the proposed oAPT workflow in clinical implementation. However, since the oAPT workflow essentially had the same components as the widely-used offline ART workflow, oAPT could theoretically be applied to a wide range of clinical scenarios with small and large anatomical changes in various disease sites such as lung, esophagus, and head and neck cancer.

During the new structure DICOM file generation, the use of the AI tool, AccuContour$^{TM}$, was conservatively restricted to the delineation of OARs only, while the targets were still propagated by RIR with possible manual modification. Given the continuous efforts dedicated to exploit AccuContour$^{TM}$ in the delineation of targets[38,39], it is possible to train institution-specific CTV auto-segmentation models through transfer learning. As a result, then other different targets can be generated by automatic margin expansion for subsequent clinical workflow.

In the proposed oAPT workflow, the verification plan was generated by a forward dose calculation on the *vCT* using the most recent previously approved clinical plan. Alternative strategy to verify the plan quality on the *vCT* is to derive the accumulated dose[14,17] from all previously approved plans on the *vCT* that triggers the need for DIR to deform the dose, either using the conventional algorithms[44-46] or the AI-based model[47,48]. The forward dose calculation method placed more emphasis on the influence caused by anatomical changes in the current fraction, thus was more conservative and led to increased probablity of re-optimization. In contrast, the accumulated dose calculation focused more on the impact of anatomical change on the overall plan dose distribution from all the previously delivered fractions and the current fraction, thus, in part, alleviated the impact on the plan dose distribution in the current fraction and reduced the possibility for re-optimization. However, it introduces other concerns from the uncertainties in dose deformation due to DIR. As a result, the forward dose calculation method was adopted for the verification plan in the proposed oAPT workflow.

There are several limitations in this study. First, the time to obtain *vCT* using CT-on-rails was not included in the oAPT time estimation shown in Table 3. In our practice, it took on average about 4 minutes, including the times for moving patient from setup poisiton to the CT-on-rails, taking CT scan, and moving patient back to the setup position. An the *vCTs* currently used had more slices than necessary, which not only prolonged the CT scanning time, but also the subsequential image processing time. We have a new protocol for scanning a much focused area that can significantly reduce the time. Once we gather enough patients using the new CT scanning protocol in the future, we will re-do efficacy validation and time cost evaluation for the oAPT workflow. Second, the current oAPT workflow might suffer uncertainties arising from the RIR in the verification plan. However, such uncertainties would not have any influence on the re-optimization since the re-plans were generated from scratch using the new initial spot arrangement based on *vCT*. Third, in the initial spot arrangement, the beam angles were still inherited from the initial plan, i.e., T180G95 and T0G95 in this study. The function of beam angle selection either through manual selection or beam angle optimization[49,50] will be incoporated into the proposed oAPT workflow in the future.

**Conclusion**

An AI-faciliated MC-based oAPT workflow for PBSPT was developed with new functionalities including AI-based auto-segmentation, initial spot arrangement, and LET-based biological effect evaluation. This feasibility study of clinical implementation using eleven prostate patients confirmed the effectiveness of the proposed oAPT workflow in the IMPT treatment of prostate cancer patients.